\documentclass[a4paper,12pt]{article} \usepackage{amssymb} \usepackage{amsmath}

\usepackage{graphics}

\usepackage{latexsym}

 \title{Integrable Hierarchy of Multi-Component Kaup -Boussinesq Equations}

\vspace{4cm}

\author {Metin G{\" u}rses\thanks{Email:gurses@fen.bilkent.edu.tr}\\ {\small Department of Mathematics, Faculty of Science} \\ {\small Bilkent
University, 06800 Ankara - Turkey}\\ }

   \date{\nonumber} 
\begin{document}
\maketitle \date{\nonumber}

\baselineskip 17pt

\numberwithin{equation}{section}

\vspace{3cm}

\begin{abstract}
By using the Lax approach we find the integrable hierarchy  of the
two and three field Kaup-Boussinesq equations.  We then give a
multi-component Kaup-Boussinesq equations and their recursion
operators. Finally we show that all multi-component
Kaup-Boussinesq equations are the degenerate Svinolupov KdV
systems.
\end{abstract}

\newtheorem{thm}{Theorem}[section] \newtheorem{Le}{Lemma}[section] \newtheorem{defi}{Definition}[section]

\def \part {\partial}
\def \be {\begin{equation}}
\def \ee {\end{equation}}
\def \bea {\begin{eqnarray}}
\def \eea {\end{eqnarray}}
\def \ba{\begin{array}}
\def \ea {\end{array}}
\def \si {\sigma}
\def \al {\alpha}
\def \la {\lambda}
\def \D {\displaystyle}

\newpage

\section{Introduction}

Systems of nonlinear evolution equations which are integrable are
not so many. Well-known ones are the coupled nonlinear Schrodinger
equations on simple Lie algebras \cite{fordy-kul}, \cite{fordy},
coupled KdV equations on Jordan algebras \cite{svin1},
\cite{svin3}, and coupled KdV, mKdV and Nonlinear Schrodinger
equations and their super extensions on symmetric and homogeneous
spaces \cite{gur0}, \cite{gur01}. The way one obtains such systems
is either by starting with soliton connections on the
corresponding algebras \cite{gur0}-\cite{gerd2} and imposing the
zero-curvature condition or by assuming the form of the recursion
operator of the corresponding system \cite{gur1}, \cite{gur2}.
Most cases these two approaches are in agreement \cite{gur-sok}.
This means that one imply the other.

In this work we start with a Lax representation of the KB
equations and then derive the whole hierarchy and the recursion
operator. This approach is similar to the one to obtain the
hierarchy of the KdV equation \cite{lax}, \cite{abl}. The Lax pair
for the KdV equation $u_{t}=-u_{xxx}+6uu_{x}$ is given by

\begin{eqnarray}
\psi_{xx}&=&(-\lambda+u(t,x))\, \psi, \label{lax}\\
\psi_{t}&=&A\, \psi-\frac{1}{2} A_{x}\, \psi
\end{eqnarray}
where $A=2u+4\lambda$ for the KdV equation and it is a polynomial
of the spectral parameter $\lambda$ for the integrable hierarchy
of the KdV equation. One can first find the recursion operator
${\bf R}=-D^2+4u+2u_{x} D^{-1}$ and then the whole hierarchy of
the KdV equation. For the multi-component KB equations the Lax
operator in (\ref{lax}) is also a second order differential
operator but it is a polynomial of degree $\ell$ in the spectral
parameter $\lambda$.

By using the above Lax approach for the KdV equation we will give
a new integrable system of evolution equations named as
multi-component Kaup-Boussinesq (KB) equations. We shall do this
starting from two and three field KB equations. Two component
Kaup-Boussinesq equation has been studied earlier by several
authors \cite{kaup}-\cite{ivan}. We introduce two component KB
equations in sections 2 and 3. In these sections we give the full
hierarchy of two field KB equations with their recursion operator.
In section 4 we give the three component extension of KB equations
with the corresponding recursion operator. Then, in section 5 we
give the multi-component KB equations with the recursion operator.
It is shown in section 6 that the system of multi-component KB
equations is a sub-class of degenerate Svinolupov KdV system.

\section{Two Field KB Equations}

Recently Ivanov and Lyons \cite{ivan} introduced a Lax pair

\begin{eqnarray} \psi_{xx}&=&[-\lambda^2+\lambda u(x,t)+\frac{k}{2} u^2(x,t)+v(x,t)]\, \psi,
\\ \psi_{t}&=&-(\lambda+\frac{1}{2}u(x,t))\,
\psi_{x}+\frac{1}{4} u_{x} \psi,
\end{eqnarray}
leading to the coupled evolution equations
\begin{eqnarray} &&u_{t}+v_{x}+(\frac{3}{2}+k) u u_{x}=0,
\label{eq0}\\
&&v_{t}-\frac{1}{4} u_{xxx}+(u v)_{x}-(\frac{1}{2}+k)u v_{x}-k (\frac{1}{2}+k)u^2 u_{x}=0 \label{eq1}
\end{eqnarray}
where $\lambda$ is
the spectral parameter and $k$ is an arbitrary constant.

We shall now find the hierarchy of the above equations just like determining the hierarchy of the KdV equation.
Let the Lax operator ${\bf L}$ be
given by

\begin{equation}\label{lax0}
{\bf L}=D^2-\lambda u+\frac{k}{2} u^2+v
\end{equation}
Then the first part of the Lax equation is the
eigen-value equation for $L$
\begin{equation}\label{lax1} {\bf L}\, \psi=-\lambda^2\, \psi
\end{equation}
Let the time evolution of $\psi$ be given by
\begin{equation}\label{lax2} \psi_{t}=A \psi_{x}+B \psi,
\end{equation}
where $A$ and $B$ depend on the dependent variables $u$, $v$ and their
derivatives with respect to $x$ and also on the spectral parameter $\lambda$.
Compatibility of the Lax equations (\ref{lax1}) and (\ref{lax2}) gives
$B=-\frac{1}{2} A_{x}$ and

\begin{eqnarray}\label{eq2} &A_{xxx}+[-2k u^2 +4\lambda^2 -4 \lambda u -4v] A_{x}-[2k u u_{x}+  \nonumber \\
&2\lambda u_{x}+ 2v_{x}] A+2 k u u_{t}+2 \lambda u_{t}+2v_{t}=0
\end{eqnarray}
One can solve this equation by assuming $A$ as a polynomial of $\lambda$
(analytic in $\lambda$). For instance, by assuming $A$ as a third order polynomial in $\lambda$, a solution of this equation is obtained as

\begin{eqnarray} &A=a_{0}\,[16 \lambda^3+8 \lambda^2 u+2 \lambda (2ku^2+3u^2+4v)-2u_{xx}+6ku^3+5u^3+12uv] \\
&+a_{1}\,[16 \lambda^2+8 \lambda u+2 (2k u^2+3u^2+4v)]+a_{2}\, [4
\lambda+2u]+a_{3}
\end{eqnarray}
where $a_{0}$, $a_{1}$, $a_{2}$ and $a_{3}$ are arbitrary constant. The corresponding
evolution equations are given as follows:

\begin{eqnarray} u_{t}&=&\frac{a_{0}}{16}\,[-(4ku+10u)u_{xxx}-12(k+1)u_{x}u_{xx}+(12k^2+60k+35)u^3 u_{x} \nonumber \\
&&+(24k+60)uvu_{x}+(12k+30)u^2 v_{x}+24vv_{x}]+\frac{a_{1}}{16}\,[-4u_{xxx}+ \nonumber \\
&&(36k+30)u^2u_{x} +24 (uv)_{x}]+\frac{a_{2}}{16}\,[2(2k+3)uu_{x}+4
v_{x}]-\frac{a_{3}}{16}\,u_{x}
\end{eqnarray}

\begin{eqnarray} v_{t}&=&\frac{a_{0}}{32}\,[2u_{xxxxx}+(8k^2-2k-15)u^2 u_{xxx}-
20vu_{xxx}+(24k^2-72k-90)u u_{x} \nonumber \\
&&u_{xx}-40v_{x}u_{xx}-(36k+30)(u_{x})^3-36u_{x}v_{xx}-(24k^3+72k^2+30k)u^4 u_{x} \nonumber \\
&&+(60-48k^2)v u^2 u_{x}+48v^2 u_{x}+(8k-12)u
v_{xxx}-(24k^2+24k-10)u^3 v_{x} \nonumber \\
&&+(-48k+72)uv v_{x}]+\frac{a_{1}}{32}\,[-12uu_{xxx}-(36+24k)u_{x}u_{xx} \nonumber \\
&&-24k(2k+1)u^3 u_{x}+48uvu_{x}-8v_{xxx}+12(1-2k)u^2v_{x}+48vv_{x}] \nonumber\\
&&+\frac{a_{2}}{32}\,[-2u_{xxx}-4k(2k+1)u^2u_{x}+8vu_{x}+4(-2k+1)uv_{x}]-\frac{a_{3}}{16}\,v_{x}
\end{eqnarray}

The evolution equations presented in (\ref{eq0}) and (\ref{eq1})
correspond to to the choices $a_{0}=a_{1}=a_{3}=0$ and $a_{2}=-4$.
Then we can find the first four members of the hierarchy as
follows

\vspace{0.3cm} \noindent {\bf N=0}: All constants $a_{i}=0~
(i=0,1,2)$ in the above equations except $a_{3}$.

\begin{eqnarray} u_{t_{0}}+u_{x}=0, \\ v_{t_{0}}+v_{x}=0, \end{eqnarray}

\vspace{0.3cm} \noindent {\bf N=1}: All constants
$a_{i}=0~(i=0,1,3)$ in the above equations except $a_{2}$.

\begin{eqnarray}
&&u_{t_{1}}+v_{x}+(\frac{3}{2}+k) u u_{x}=0, \\
&&v_{t_{1}}-\frac{1}{4} u_{xxx}+(u v)_{x}-(\frac{1}{2}+k)u v_{x}-k (\frac{1}{2}+k)u^2
u_{x}=0
\end{eqnarray}

\vspace{0.3cm} \noindent {\bf N=2}: All constants
$a_{i}=0~(i=0,2,3)$ in the above equations except $a_{1}$.

\begin{eqnarray} &&u_{t_{2}}=\frac{a_{1}}{32}\,[-12uu_{xxx}-(36+24k)u_{x}u_{xx} \nonumber \\
&&-24k(2k+1)u^3
u_{x}+48uvu_{x}-8v_{xxx}+12(1-2k)u^2v_{x}+48vv_{x}],\\
&&v_{t_{2}}=\frac{a_{1}}{32}\,[-12uu_{xxx}-(36+24k)u_{x}u_{xx} \nonumber \\ &&-24k(2k+1)u^3
u_{x}+48uvu_{x}-8v_{xxx}+12(1-2k)u^2v_{x}+48vv_{x}]
\end{eqnarray}

\vspace{0.3cm} \noindent {\bf N=3}: All constants
$a_{i}=0~(i=1,2,3)$ in the above equations except $a_{0}$.

\begin{eqnarray}
&&u_{t_{3}}=\frac{a_{0}}{16}\,[-(4ku+10u)u_{xxx}-12(k+1)u_{x}u_{xx}+(12k^2+60k+35)u^3 u_{x} \nonumber \\
&&+(24k+60)uvu_{x}+(12k+30)u^2 v_{x}+24vv_{x}],\\
&&v_{t_{3}}=\frac{a_{0}}{32}\,[2u_{xxxxx}+(8k^2-2k-15)u^2 u_{xxx}-20vu_{xxx}+(24k^2-72k-90)u u_{x}
\nonumber \\
&&u_{xx}-40v_{x}u_{xx}-(36k+30)(u_{x})^3-36u_{x}v_{xx}-(24k^3+72k^2+30k)u^4 u_{x} \nonumber \\
&&+(60-48k^2)v u^2 u_{x}+48v^2
u_{x}+(8k-12)u v_{xxx}-(24k^2+24k-10)u^3 v_{x} \nonumber\\
&&+(-48k+72)uv v_{x}]
\end{eqnarray}

By defining a new variable $\bar{v}=v+\frac{k}{2}\, u^2$ it is possible to eliminate $k$ dependence
of the equations (\ref{eq0}) and (\ref{eq1}). The
new equations are

\begin{eqnarray}
&&u_{t}+\bar{v}_{x}+\frac{3}{2} u u_{x}=0, \\
&&v_{t}-\frac{1}{4} u_{xxx}+ \bar{v} u_{x}+\frac{1}{2}u
\bar{v}_{x}=0.
\end{eqnarray}
 Although Eqs.(\ref{eq0}) and (\ref{eq1}) are equivalent, for any vale of $k$, to the
 above system of evolution equations corresponding to $k=0$, we will keep them in the
 sequel because the case with $k=-\frac{1}{2}$ corresponds to the
 Kaup-Boussinesq (KB) equations \cite{kaup}

\begin{eqnarray}
&&u_{t}+v_{x}+u u_{x}=0, \\
&&v_{t}-\frac{1}{4} u_{xxx}+  (uv)_{x}=0.
\end{eqnarray}
KB equation later studied by several authors \cite{zak},
\cite{pav}. We shall call the system of evolution equations
obtained here  as generalized KB equations. In section 3 we shall
find $\ell=3$ (three dependent variables) system of KB equations.
In section 4 we shall show that, for any $\ell$, the KB systems
turn out to be the Fokas-Liu extension of the Svinolupov KdV
systems  \cite{gur1}, \cite{gur2}.

\section{Hierarchy and the Recursion Operator of Two Field KB Equations}
The first three members of the hierarchy are given in the previous
section. Here in this section we shall determine the full
hierarchy by computing the recursion operator of the three field
KB equations. Let the function $A$ in (\ref{eq2}) be an analytic
function of $\lambda$, then
\begin{equation}
A=\sum_{n=0}^{N}\,A_{n}\, \lambda^{N-n}
\end{equation}
where $A_{n}$'s are functions of $u,v$ and their
$x$ partial derivatives. We obtain
\begin{eqnarray}
&&2u_{t_{N}}-2(2uD+u_{x}) A_{N}+{\bf M}_{2}
A_{N-1}=0, \label{eq3}\\
&&2v_{t_{N}}+[{\bf M}_{2}+2ku(2uD+u_{x})
A_{N}-k u \,{\bf M} A_{N-1}=0, \label{eq4}
\end{eqnarray}
and the
recursion relations among $A_{m}$, ($0 \le m \le N-2$).

\begin{equation}\label{eq5}
{\bf M}_{2} A_{m}+4
A_{m+2},x-2(2uD+u_{x}) A_{m+1}=0,~~0 \le m \le N-2
\end{equation}
with $A_{0}=a_{0}=$ constant and $A_{1}=\frac{1}{2}a_{0}u$ and so on.
In the above expressions $D$ is the differential operator with
respect $x$ and

\begin{equation} {\bf M}_{2}=D^3-2(2v+ku^2)D-2 (kuu_{x}+v_{x})
\end{equation}

Using (\ref{eq3}), (\ref{eq4}) and the recursion relations (\ref{eq5}) we obtain that

\begin{equation}
 {u_{t_{N}} \choose v_{t_{N}}}={\bf R}{u_{t_{N-1}} \choose
 v_{t_{N-1}}},
\end{equation} where ${\bf R}$ is the recursion operator of the hierarchy

\begin{equation}\label{rec}
{\bf R}=\begin{pmatrix}-(k+1)u
-\frac{1}{2}\,u_{x} D^{-1}&-1 \\ \frac{1}{4}D^2-v+(\frac{1}{2}+k)ku^2-\frac{1}{2}\,v_{x}D^{-1}& ku
\end{pmatrix} \end{equation}

Then the hierarchy of the shallow water wave equations are given by

\begin{equation}
 {u_{t_{N}} \choose v_{t_{N}}}={\bf R}^{N}{u_{x} \choose
 v_{x}},~~~N=1,2,\cdots
 \end{equation}

To prove that the operator in (\ref{rec}) is the recursion operator we write
the coupled evolution equations (\ref{eq0}) and (\ref{eq1}) as

\begin{equation}
 {u_{t} \choose v_{t}}=K={-v_{x}-(\frac{3}{2}+k) u u_{x} \choose
 \frac{1}{4} u_{xxx}-(u v)_{x}+(\frac{1}{2}+k)u v_{x}+k
(\frac{1}{2}+k)u^2 u_{x} },
 \end{equation}
 Then
\begin{equation}
 {(\delta u)_{t} \choose (\delta v)_{t}}={\bf K}^{*} {\delta u \choose
\delta v },
 \end{equation}
where
\begin{equation}
{\bf K}^{*}=\begin{pmatrix} -(\frac{3}{2}+k)(u_{x}+uD)& -D \\ \frac{1}{4}D^3+k(\frac{1}{2}+k)u^2 D-v
D+2k(\frac{1}{2}+k) uu_{x}-(\frac{1}{2}-k)v_{x}& -(\frac{1}{2}-k)u D-u_{x}
\end{pmatrix} \end{equation}
Then it is easy to show that the recursion operator for symmetries
satisfies (see \cite{olv}, \cite{blaz})

\begin{equation}\label{receqn}
{\bf R}_{t}=[{\bf K}^{*}, {\bf R}]
\end{equation} We have the Hamilton operators

\begin{equation}
\theta_{0}=\begin{pmatrix}0& D \\ D& 0
\end{pmatrix}
\end{equation}

\begin{equation}
\theta_{1}=\begin{pmatrix}-2D& -2(k+1)uD -u_{x}\\
2ukD & \frac{1}{2}D^3+(-2v+(1+2k)ku^2)D-v_{x}
\end{pmatrix}
\end{equation}
so that ${\bf R}=\theta_{1}\, \theta_{0}^{-1}$.

\section{Three field KB Equations }

Let the Lax operator ${\bf L}$ be given by
\begin{equation}\label{lax3}
{\bf L}=D^2-\lambda^2 u+\lambda v +w
\end{equation}
where $u$, $v$, and $w$
are functions of $x$ and $t$. Then the first part of the Lax equation is the eigen-value equation for $L$
\begin{equation}\label{lax4}
{\bf L}\,
\psi=-\lambda^3\, \psi
\end{equation}
Similarly let the time evolution of $\psi$ be given by

\begin{equation}\label{lax5}
\psi_{t}=A \psi_{x}+B \psi,
\end{equation}
where $A$ and $B$ depend on the dependent variables $u$, $v$ and their derivatives with respect
to $x$ and also on the spectral
parameter $\lambda$. Compatibility of the Lax equations (\ref{lax4}) and (\ref{lax5}) gives $B=-\frac{1}{2} A_{x}$ and

\begin{eqnarray}\label{eq2}
&A_{xxx}+[4\lambda^3 -4 \lambda^2 u -4 \lambda v-4w] A_{x}-[2\lambda^2 u_{x}+ 2 \lambda v_{x} \nonumber \\
&+2w_{x}] A+2 \lambda^2 u_{t}+2 \lambda
v_{t}+2 w_{t}=0 \end{eqnarray}
One can solve this equation by assuming $A$ as a polynomial of $\lambda$ (analytic in $\lambda$). For instance, by
assuming $A$ as a third order polynomial in $\lambda$, a solution of this equation is obtained as
\begin{eqnarray} &A=(a_{0}\,[128\lambda^4+64
\lambda^3 u+16 \lambda^2 (3u^2+4v)+8 \lambda (5 u^3+12 uv +8w)-16 u_{xx} \nonumber \\
&+35u^4+120u^2 v+96uw+8w]+a_{1}\,[128 \lambda^3+ 64 \lambda^2 u+
16 \lambda (3u^2+4v)+8(5u^3+12uv+8w)] \nonumber \\
&+a_{2}\, [128 \lambda^2+64 \lambda u +16(3u^2+4v)]+a_{3}[128 \lambda+64u]+128 a4)/128
\end{eqnarray}
where $a_{0}$, $a_{1}$, $a_{2}$,$a_{3}$ and $a_{4}$ are arbitrary constant.
The corresponding evolution equations are given as
follows:

\vspace{0.3cm} \noindent
{\bf N=1}. $a_{0}=a_{1}=a_{2}=a_{4}=0$
and $a_{3}=1$.

\begin{eqnarray}
 u_{t}&=&\frac{3}{2} uu_{x}+v_{x}, \\
 v_{t}&=&vu_{x}+\frac{1}{2}\,uv_{x}+w_{x},\\
w_{t}&=&-\frac{1}{4} u_{xxx}+wu_{x}+\frac{1}{2}uw_{x}
\end{eqnarray}

\vspace{0.3cm}
\noindent {\bf N=2}. $a_{0}=a_{1}=a_{3}=a_{4}=0$
and $a_{2}=1$.

\begin{eqnarray}
u_{t}&= &\frac{15}{8} u^2 u_{x}+\frac{3}{2} (uv)_{x}+w_{x}, \\
 v_{t}&=&-\frac{1}{4}\,u_{xxx}+\frac{3}{2}\,u v u_{x}+\frac{3}{8}\,3u^2
v_{x}+wu_{x}+\frac{3}{2}\,v v_{x}+\frac{1}{2}\,u w_{x},\\
 w_{t}&=&-\frac{3}{8} u u_{xxx}-\frac{1}{4} v_{xxx}-\frac{9}{8} u_{x}u_{xx}+\frac{3}{2}\,u w u_{x}+wv_{x}+
\frac{3}{8}u^2w_{x}+\frac{1}{2} v w_{x},
\end{eqnarray}
etc. To find  all members the of hierarchy of three field
evolution equations we let

\begin{equation} A=\sum_{n=0}^{N}\,A_{n}\, \lambda^{N-n} \end{equation} where $A_{n}$'s are functions of $u,v$ and $w$ and their $x$ partial
derivatives. Then we obtain

\begin{equation} A_{0}=a_{0},~~ A_{1}=\frac{a_{0}}{2} u+a_{1},~~~A_{2}=\frac{3a_{0}}{8} u^2+\frac{1}{2}a_{1} u
+\frac{1}{2}a_{0} v+a_{2}
\end{equation}
where $a_{0}$,$a_{1}$ and $a_{2}$ are arbitrary constants. The evolution equations are given as follows

\begin{eqnarray} & 2u_{t_{N}}+{\bf M}_{3}\,
A_{N-2}-2(2uD+u_{x})\,A_{N}-2(2vD+v_{x}) A_{N-1}=0, \\
&2v_{t_{N}}+{\bf M}_{3}\,A_{N-1}-2(2vD+v_{x}) A_{N}=0,\\
&2w_{t_{N}}+{\bf M}_{3} A_{N}=0
\end{eqnarray}
with the recursion relations
\begin{equation} {\bf M}_{3} \, A_{n}+4 A_{n+3,\, x}-2(2uD+u_{x})\, A_{n+2}-2 (2vD+v_{x}) a_{n+1}=0
\end{equation}
where $0 \le n \le N-3$ and
\begin{equation} {\bf M}_{3}=D^3-4wD-2w_{x}
\end{equation}

It is straight forward to show

\begin{equation}
 \begin{pmatrix} u_{t_{N}} \\ v_{t_{N}} \\w_{t_{N}}
 \end{pmatrix}={\bf R}\begin{pmatrix} u_{t_{N-1}} \\ v_{t_{N-1}}
 \\w_{t_{N-1}}.
 \end{pmatrix}, ~~N=1,2,\cdots
\end{equation} where ${\bf R}$ is the recursion operator

\begin{equation}\label{rec1}
{\bf R}=\begin{pmatrix} (u D +\frac{1}{2}\,u_{x}) D^{-1}~~& ~~1 ~~&0\\ (v D +\frac{1}{2}\,v_{x}) D^{-1}~~&~~0~~&1\\
 -\frac{1}{4}{\bf
M}_{3}D^{-1}~~&~~ 0~~&0
\end{pmatrix} \end{equation}
One can verify easily that ${\bf R}$ satisfies the equation
(\ref{receqn}) where ${\bf K}^{*}$ is given by

\begin{equation}\label{K12}
{\bf K}^{*}=\begin{pmatrix} \frac{3}{2}u D +\frac{3}{2}\,u_{x}~~& ~~D ~~&0\\
v D +\frac{1}{2}\,v_{x}~~&~~\frac{1}{2}u D +\,u_{x}~& D\\
 -\frac{1}{4}D^3+w D+\frac{1}{2}w_{x}~~&~~ 0~~&\frac{1}{2}u D +\,u_{x}
\end{pmatrix} \end{equation}

Then the hierarchy of the three field KB is given by

\begin{equation}
\begin{pmatrix} u_{t_{N}} \\ v_{t_{N}} \\w_{t_{N}}
\end{pmatrix} ={\bf R}^{N}
\begin{pmatrix} u_{x} \\
 v_{x}\\ w_{x}
 \end{pmatrix} ,~~~N=1,2,\cdots
 \end{equation}
$N=1$ gives the three field KB Equations and all $N >1$ cases are
generalized symmetries of the three field KB equations.

\section{Multi-Component KB Equations}

Multi-component KB equations can be obtained from the Lax operator

\begin{equation}
{\bf L}=D^2-\sum_{k=1}^{\ell}\, \lambda^{k-1}\,q^{k}(x,t)
\end{equation}
where $q^{k}(x,t)$, $(k=1,2, \cdots, \ell)$ are the multi-KB
fields, as we did in the earlier sections. Here $\ell$ is a
positive integer grater or equal to two. The Lax equations in this
case take the form

\begin{eqnarray}
{\bf L}\, \psi &=&-\lambda^{\ell}\, \psi, \\
\psi_{t}&=&A\, \psi_{x}-\frac{1}{2}A_{x}\, \psi,
\end{eqnarray}
where $A$ is a polynomial of the spectral parameter $\ell$. To
obtain the multi-component KB equations and their recursion
operators by the Lax operator given above for arbitrary positive
integer $\ell$ is very lengthy. Instead we shall make use our
experience from the two-field and three field KB equations and
their recursion operators in sections 2-4.  The multi system of KB
equations are given as follows

\begin{eqnarray}
u_{t}&=&\frac{3}{2}uu_{x}+q^{2}_{x}, \label{multi1}\\
q^{2}_{t}&=&q^{2} u_{x}+\frac{1}{2} u q^{2}_{x}+q^{3}_{x}, \\
q^{3}_{t}&=&q^{3}
u_{x}+\frac{1}{2} u q^{3}_{x}+q^{4}_{x}, \\
\vdots &~~&\vdots~~~ ~\vdots ~~~~ \vdots ~~~~ \vdots\\ q^{\ell-1}_{t}&=&q^{\ell-1} u_{x}+\frac{1}{2} u
q^{\ell-1}_{x}+w_{x}, \\
w_{t}&=&-\frac{1}{4} u_{xxx}+w u_{x}+\frac{1}{2} u w_{x}, \label{multison}
\end{eqnarray}
where we took $q^{1}=u$ and
$q^{\ell}=w$. The recursion operator of the system can be given by
\begin{equation}\label{multi-rec}
{\bf R}=\begin{pmatrix} u+\frac{1}{2} u_{x}
D^{-1} & 1 &0 &0 &\ldots &0\\ q^{2}+\frac{1}{2} q^{2}_{x} D^{-1}& 0& 1&0& \dots &0 \\
q^{3}+\frac{1}{2} q^{3}_{x} D^{-1}& 0& 0&1& \dots &0 \\ \vdots &
\vdots & \vdots & \vdots & \ddots &\vdots \\ q^{\ell-1}+\frac{1}{2} q^{\ell-1}_{x} D^{-1}& 0& 0&0& \dots &1 \\
-\frac{1}{4} M_{\ell} D^{-1} &0&0&0
&\ldots& 0 \end{pmatrix}
\end{equation}
where
\begin{equation}
 {\bf M}_{\ell}=D^3-4wD-2w_{x}
 \end{equation}

It is not difficult to show that the KB equations (\ref{multi1})-(\ref{multison}) are integrable
 and the operator in (\ref{multi-rec}) is the
recursion operator of the system.

One can show that the recursion operator ${\bf R}$ satisfies the equation (\ref{receqn}) where

\begin{equation}\label{multi-kstar}
{\bf K}^{*}= \begin{pmatrix} \frac{3}{2}u_{x}+\frac{3}{2} u D & D &0 &0 &\ldots &0&0\\ \frac{1}{2} q^{2}_{x}+q^{2}
D& u_{x}+\frac{1}{2} u D& D&0& \dots &0 &0\\ \frac{1}{2}q^{3}_{x}+q^{3} D& 0&u_{x}+\frac{1}{2} u D& D& \dots &0&0 \\
\vdots & \vdots & \vdots & \vdots &
\ddots &\vdots &\vdots\\ \frac{1}{2}q^{\ell-1}_{x}+q^{\ell-1}D&0&0 &0&\dots &u_{x}+\frac{1}{2} u D& D \\
 -\frac{1}{4} D^3+\frac{1}{2} w_{x}+wD
&0&0 &0&\ldots&0& u_{x}+\frac{1}{2} u D
\end{pmatrix}. \end{equation}
Hence  the KB equations (\ref{multi1})-(\ref{multison}) are integrable and the
operator in (\ref{multi-rec}) is the recursion operator of the system.

\section{Multi-Component KB Equations as Svinolupov KdV System}

The system of KB  equations (\ref{multi1})-(\ref{multison}) and
the corresponding recursion operator (\ref{multi-rec}) look like
 the coupled integrable KdV systems studied earlier
\cite{gur1}, \cite{gur2}. Let $q^{i}=(q^{1},q^{2}, \cdots ,
q^{\ell})$ be $\ell$-number of fields all depending on $x$ and
$t$. Let $q^{i}$ with $i=1,2, \cdots, \ell$ satisfy the following
coupled evolution equations

\begin{equation}\label{coupled} q^{i}_{t}=b^{i}\,_{k}\,
q^{k}+S^{i}\,_{jk}\,q^{j}\,q^{k}_{x}+\chi^{i}\,_{k}\,q^{k}_{x}, ~~i=1,2,\cdots, \ell
\end{equation}
where all repeated indices are summed up from 1 to
$\ell$. The coupled evolution equations (\ref{coupled}) are integrable if the
constants $b^{i}\,_{k}$, $S^{i}\,_{jk}$ and $\chi^{i}\,_{k}$ satisfy
certain conditions \cite{gur1}, \cite{gur2}. If $b^{i}\,_{j}=\delta^{i}_{j}$ and
$\chi^{i}\,_{j}=0$ and $S^{i}\,_{jk}$ are the structure constants of
a Jordan algebra then the system of equations (\ref{coupled}) are
integrable \cite{svin1}-\cite{svin3}. All other cases were studied in \cite{gur1},
\cite{gur2}. In particular if $det(b)=0$ and $\chi^{i}\,_{j}=0$ such
systems are called as degenerate Svinolupov KdV systems. The case when $det(b)=0$
and $\chi^{i}\,_{j} \ne 0$ is known as the Fokas-Liu extension of the
degenerate Svinolupov KdV systems \cite{fok}, \cite{gur2}.  For the integrable
cases the corresponding recursion operators are given by

\begin{equation}\label{kdv-rec}
{\bf R}^{i}\,_{j}=b^{i}\,_{j}\, D^2+A^{i}\,_{jk}\,q^{k}+C^{i}\,_{jk}\, q^{k}_{x}\,D^{-1}+ w^{i}\,_{j},~~
i,j=1,2,\cdots, \ell
\end{equation}
where $A^{i}\,_{jk}$, $C^{i}\,_{jk}$, $w^{i}\,_{j}$ are given in terms of $S^{i}\,_{jk}$ and $\chi^{i}\,_{j}$,
\cite{gur1}, \cite{gur2}. The form of the recursion operator is valid for all cases discussed above.

The multi-components KB systems fall into the  Fokas-Liu extension of the degenerate Svinolupov KdV system
where the corresponding constants are given
by

\begin{eqnarray}
&b^{i}\,_{j}=-\frac{1}{4}\, \delta^{i}_{\ell}\,\delta^{1}_{j},~~\chi^{i}\,_{j}=w^{i}\,_{j}=\delta^{i}\,_{j-1},\\
&A^{i}\,_{jk}=\delta^{1}\,_{j}\,\delta^{i}\,_{k},~~B^{i}\,_{jk}=\frac{1}{2}\,A^{i}\,_{jk},\\
&S^{i}\,_{jk}=\delta^{i}\,_{j}\,
\delta^{1}\,_{k}+\frac{1}{2}\, \delta_{i}\,_{k}\, \delta^{1}\,_{j}
\end{eqnarray}
where $\delta^{i}_{j}$ is the Kronecker $\delta$-symbol and $det(
b)=0$. Then the multi-component KB equations
(\ref{multi1})-(\ref{multison}) take the form

\begin{equation}\label{coupled1}
q^{i}_{t}=-\frac{1}{4}\delta^{i}_{\ell}\,u_{xxx}\, +q^{i}\,u_{x}+ \frac{1}{2}\,u \,q^{i}_{x}+q^{i+1}_{x},
~~i=1,2,\cdots, \ell
\end{equation} where we took  $q^{\ell+1}=0$.

The hierarchy of multi-component KB equations is given by

\begin{equation}\label{coupled2}
q^{i}_{t_{N}}=({\bf R}^{N})^{i}\,_{k}\,q^{k}_{x} ~~i=1,2,\cdots, \ell
\end{equation}
where $N=1,2, \cdots $. The case $N=1$ is the multi-component KB
equations (\ref{coupled1}). All others $N > 1$ correspond the
integrable family or higher generalized symmetries of
multi-component KB equations. Components of the $\ell \times \ell$
matrix recursion operator (\ref{multi-rec}) which is compatible
with (\ref{kdv-rec}), in index notation, is given by

\begin{equation}\label{rec4}
{\bf R}^{i}\,_{j}=-\frac{1}{4}\, \delta^{i}_{\ell}\,\delta^{1}_{j}\, D^2+ \delta^{1}_{j}\,q^{i}+
\frac{1}{2}\,\delta^{1}_{j}\,
q^{i}_{x}\,D^{-1}+ \delta^{i}_{j-1},~~ i,j=1,2,\cdots, \ell
\end{equation}
with

\begin{equation}\label{k-index}
({\bf K^{*}})^{i}\,_{j}=-\frac{1}{4}\,
\delta^{i}_{\ell}\,\delta^{1}_{j}\, D^3+
\delta^{1}_{j}\,(q^{i}\,D+\frac{1}{2} q^{i}_{x})+
\delta^{i}_{j}\,(u_{x}+\frac{1}{2} u \,D)+ \delta^{i}_{j-1} \,D ~~
i,j=1,2,\cdots, \ell
\end{equation}
The evolution equation (\ref{receqn}) for the
recursion operator for arbitrary $\ell$ in index notation is given by

\begin{equation}
\frac{d}{dt}\, {\bf R}^{i}_{j}=({\bf K}^{*})^{i}_{k}\, {\bf R}^{k}_{j}-{\bf R}^{i}_{k}\, ({\bf K}^{*})^{k}_{j}
~~i,j=1,2,\cdots, \ell
\end{equation}
It is easy to verify this equation with ${\bf R}$ and ${\bf
K}^{*}$  given in (\ref{rec4}) and (\ref{k-index}) respectively
for arbitrary $\ell$ provided that $q^{i}$'s satisfy the
 multi-component KB equations (\ref{coupled1}).

\section{Concluding Remarks}

~~~   We found the hierarchy (generalized symmetries) of the
integrable two field and three field Kaup-Boussinesq equations.
Integrability of these equations are supported by giving the
corresponding recursion and compatible Hamilton operators.

We presented a new multi-component (arbitrary number of fields)
generalization of the Kaup-Boussinesq equations. This system of
equations is integrable. It has both a Lax representation and a
symmetry recursion operator. It is also noticed that this system
is a special case of the Fokas-Liu extension of the degenerate
Svinolupov KdV system. Conservation laws and the bi-Hamiltonian
structure of the multi-component KB equations will be communicated
in a subsequent work.

\vspace{2cm}

\noindent {\bf Acknowledgment}:
 This work is partially supported by
the Scientific and Technological Research Council of Turkey.

\end{document}